\documentclass[aps,prd,12pt,notitlepage,nofootinbib,tightenlines]{revtex4-1}
\usepackage{amsmath}
\usepackage{amssymb}
\usepackage{epsfig}
\usepackage{color,graphicx}
\usepackage{bm}
\usepackage{times}
\usepackage{braket}
\usepackage{slashed}
\usepackage{hyperref}

\newcommand{\beq}{\begin{eqnarray}}
\newcommand{\eeq}{\end{eqnarray}}

\newcommand{\non}{\nonumber\\ }

\newcommand{\etap}{\eta^{(\prime)}}
\newcommand{\psl}{ p \hspace{-2.0truemm}/ }

\newcommand{\epsl}{ \epsilon \hspace{-2.0truemm}/ }

\def \jhep{ { JHEP } }

\def \mpla{ { Mod Phys Lett A } }
\def \npb{ { Nucl Phys B} }
\def \plb{ { Phys Lett B} }
\def \pr{ { Phys Rept } }
\def \prd{ { Phys Rev D} }
\def \prl{ { Phys Rev Lett.}  }

\definecolor{Blue}{rgb}{0.,0.,1.}

\definecolor{nicegreen}{rgb}{0.1,0.5,0.1}

\hypersetup{colorlinks,citecolor=nicegreen,linkcolor=nicegreen}

\begin{document}


\title{Semileptonic decays $B \to D^{(*)} l\nu$ in the perturbative QCD
factorization approach}
\author{Ying-Ying Fan, Wen-Fei Wang, Shan Cheng, and Zhen-Jun  Xiao
\footnote{xiaozhenjun@njnu.edu.cn} }
\affiliation{ Department of Physics and Institute of Theoretical Physics,\\
Nanjing Normal University, Nanjing, Jiangsu 210023, People's Republic of China}
\date{\today}
\begin{abstract}
\noindent{\bf \large \hspace{-0.8cm} Abstract} \ \ \ \ In this paper, we study the  $B \to D^{(*)} l^- \bar{\nu}_{\rm l}$ semileptonic decays and calculate the branching ratios ${\cal B}(B \to D^{(*)} l^- \bar{\nu}_{\rm l})$
and the ratios $R(D^{(*)})$ and $R_{\rm D}^{\rm l,\tau}$ by employing the
perturbative QCD (pQCD) factorization  approach.
We find that (a) for $R(D)$ and $R(D^*)$ ratios, the pQCD predictions are
$R(D)=0.430^{+0.021}_{-0.026}$, $R(D^*)=0.301 \pm 0.013$ and agree well with BaBar's
measurements of $R(D^{(*)})$;
(b) for the newly defined $R_{\rm D}^{\rm l}$ and $R_{\rm D}^{\rm \tau}$ ratios, the pQCD
predictions are $R_{\rm D}^{\rm l} = 0.450^{+0.064}_{-0.051}$ and
$R_{\rm D}^{\rm \tau} =0.642^{+0.081}_{-0.070}$, which may be more sensitive to
the QCD dynamics of the considered semileptonic decays than $R(D^{(*)})$
and should be tested by experimental measurements.
\end{abstract}


\maketitle

{\bf Key Words} B meson semileptonic decays; The pQCD factorization approach;
           Form factors; Branching ratios


\section*{1 \hspace{0.3cm} Introduction}\label{sec:1}

The semileptonic decays $B \to D\tau\bar{\nu}_{\rm \tau}$ and
$B\to D^* \tau \bar{\nu}_{\rm \tau}$
have been previously measured by both BaBar and Belle Collaborations with
$3.8\sigma$ and $8.1\sigma$ significance \cite{babar2008,belle2007,belle2010}.
Very recently, the BaBar collaboration with their full data
greatly improved their previous analysis
and  reported their measurements for the relevant branching ratios and
the ratios $R(D^{(*)})$ of the corresponding branching ratios
\cite{prl109-101802}:
\beq
\label{eq:exp02}
{\cal R}(D) = 0.440 \pm 0.072,\quad {\cal R}(D^*) = 0.332 \pm 0.030,
\eeq
where the isospin symmetry relations ${\cal R}(D^0)={\cal R}(D^+)={\cal R}(D)$ and
${\cal R}(D^{*0})={\cal R}(D^{*+})={\cal R}(D^*)$ have been imposed,
and the statistical and systematic
uncertainties have been combined in quadrature. These BaBar results are
surprisingly larger than the standard model (SM)
predictions as given in Ref.~\cite{prd85-094025}:
\beq
{\cal R}(D)^{\rm SM} &=& 0.296\pm 0.016\; , \quad {\cal R}(D^*)^{\rm SM} = 0.252\pm 0.003,
\label{eq:sm01}
\eeq
The combined BaBar results disagree with the SM predictions by $3.4\sigma$
\cite{prl109-101802,bozek-2013}.

Since the report of BaBar measurements, this $R(D^{(*)})$ anomaly has been
studied intensively by many authors, for example, in
Refs.~\cite{prl109-071802,prd85-114502,prl109-161801,prd86-054014,jhep2013-01054,prd86-034027,
prd86-114037,mpla27-1250183,Fajfer-1301,pich-1301,R17,R18}.
Some authors treat this  $3.4\sigma$ deviation as the first evidence for new
physics (NP) in semileptonic B meson decays to $\tau$ lepton
\cite{prl109-161801,prd86-054014,jhep2013-01054,prd86-034027,prd86-114037},
such as the NP contributions from
the charged Higgs bosons in the Two-Higgs-Doublet models \cite{prd86-054014}.

Some other physicists, however, try to interpret the data in the
framework of the SM but with their own methods.
In Ref.~\cite{prl109-071802} the authors presented their SM predictions
$ {\cal R}(D)^{\rm SM} = 0.316\pm 0.014$ by using the form factors $F_{\rm 0,+}(q^{\rm 2})$
computed in unquenched lattice QCD by the
Fermilab Lattice and MILC collaborations\cite{prd85-114502}.
In Refs.~\cite{mpla27-1250183,Kosnik-1301}, furthermore, the authors
performed the same kinds of calculations by employing the relativistic quark
model\cite{mpla27-1250183} or by maximally employing the experimental
information on the relevant form factors from the data of $B \to D l\bar{\nu}_{\rm l}$
with $l=(e^-,\mu^-)$ \cite{Kosnik-1301,prl104-011802}, and found that:
\beq
{\cal R}(D)^{\rm SM} &=& 0.315\; [14],
\quad {\cal R}(D^*)^{\rm SM} = 0.260\; [14], \label{eq:mpla27} \\
{\cal R}(D)^{\rm SM} &=& 0.31\pm 0.02\; [17],
\label{eq:kosnik}
\eeq
It is easy to see that there is a clear discrepancy between these
SM predictions for $R(D^{(*)})$
\cite{prd85-094025,prl109-161801,prd85-114502,prl109-071802,mpla27-1250183,Kosnik-1301}
and the BaBar's measurements as listed in Eq.~(1).

In Refs.~\cite{prd86-114025,prd87-097501}, we studied the semileptonic
decays $B_{\rm (s)}\to(\pi,K,\eta,\etap,G) (ll,l\nu,\nu\bar{\nu})$
in the pQCD factorization approach \cite{ppnp51-85}
with the inclusion of the known  next-leading-order (NLO) contributions.
We found that all known semileptonic decays $B/B_{\rm s} \to P (ll,l\nu,\nu\bar{\nu})$
( here $P=(\pi,K,\eta,\etap,etc)$ are light pseudo-scalar mesons) can be understood
in the framework of the pQCD factorization approach\cite{prd86-114025,prd87-097501}.

Motivated by the recent BaBar's discrepancy of the measured values of
$R(D^{(*)})$ from the SM predictions, we here will calculate the branching ratios
${\cal B}(B \to D^{(*)} l^- \bar{\nu}_{\rm l})$ and the
six R(X)-ratios: the four isospin-unconstrained ratios $R(D^0)$,$R(D^{*0})$,$R(D^+)$ and
$R(D^{*+})$, as well as the two isospin-constrained ratios $R(D)$
and $R(D^*)$ in the framework of the SM by employing the pQCD approach again.
We will compare the pQCD predictions for the branching ratios and the
six $R(X)$ ratios with those as given in
Refs.~\cite{prd85-094025,prl109-161801,prd85-114502,prl109-071802,mpla27-1250183},
and the measured values of BaBar Collaboration \cite{prl109-101802}.
We also define two new ratios of the branching ratios $R_{\rm D}^{\rm l}$ and $R_{\rm D}^{\rm \tau}$,
and present the pQCD predictions for new ratios $R_{\rm D}^{\rm l,\tau}$,
which will be tested by experimental measurements.
Finally, there will be a short summary.

\section*{2 \hspace{0.3cm} Kinematics and the Wave Functions}\label{sec:2}

\begin{figure}[tbp]
\vspace{-4cm}
\centerline{\epsfxsize=16cm \epsffile{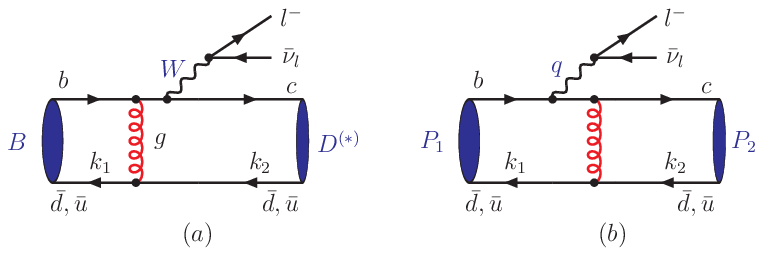}}
\vspace{-15.5cm}
\caption{ The lowest order Feynman diagrams for the semileptonic
decays $B \to D^{(*)} l^-\bar{\nu}_{\rm l}$
in the pQCD approach, the winding curves are gluons.}
\label{fig:fig1}

\end{figure}

In the pQCD approach, the lowest order Feynman diagrams for $B \to
D^{(*)} l^- \bar{\nu}_{\rm l}$ decays are displayed in Fig.\ref{fig:fig1}.
We discuss kinematics of these decays in the large-recoil (low $q^{\rm 2}$) region£¬
where
the pQCD factorization approach is applicable to the considered semileptonic
decays involving $D$ or $D^*$ as the final state meson\cite{li1995}.
In the $B$ meson rest frame, we define the $B$ meson momentum $P_{\rm 1}$,
the $D^{(*)}$ momentum $P_{\rm 2}$ in the light-cone coordinates
as\cite{prd67-054028}
\beq
\label{eq-mom-p1p2}
P_{\rm 1}=\frac{m_{\rm B}}{\sqrt{2}}(1,1,0_\bot),\quad P_{\rm 2}=\frac{r m_{\rm B}
}{\sqrt{2}} (\eta^+,\eta^-,0_\bot),
\eeq
The longitudinal polarization vector $\epsilon_{\rm L}$ and transverse polarization vector
$\epsilon_{\rm T}$ of the $D^*$ meson are given by $ \epsilon_{\rm L}= (\eta^+,-\eta^-,0_\bot)/\sqrt{2}$,
$\epsilon_{\rm T}=(0,0,1)$ with the factors $\eta^\pm = \eta \pm \sqrt{\eta^{\rm 2}-1}$ is
defined in terms of the parameter
 \beq \label{eq-eta}
  \eta =
\frac{1}{2r}\left [ 1+r^{\rm 2}-\frac{q^{\rm 2}}{m_{\rm B}^{\rm 2}}\right],
 \eeq
  where the
ratio $r=m_{\rm D}/m_{\rm B}$ or $m_{\rm D^*}/m_{\rm B}$, and $q=p_{\rm 1}-p_{\rm 2}$ is the lepton-pair momentum.
The momenta of the spectator quarks in $B$ and $D^{(*)}$ mesons are
parameterized as
\beq \label{eq-mom-k1k2}
k_{\rm 1} =(0,x_{\rm 1}\frac{m_{\rm B}}{\sqrt{2}},k_{1\bot}),\quad
k_{\rm 2}=\frac{m_{\rm B}}{\sqrt{2}}(x_{\rm 2}r\eta^+,x_{\rm 2}r\eta^-,k_{2\bot}).
\eeq

For the $B$ meson wave function, we make use of the same one as
being used for example in Refs.\cite{prd85-094003,prd86-011501,prd86-114025},
which can be written as the form of
\beq
\Phi_{\rm B}=\frac{i}{\sqrt{2N_{c}}} (\psl_{\rm B} +m_{\rm B}) \gamma_5 \phi_{\rm B} ({\bf k_{\rm 1}}).
\label{eq:bmeson}
\eeq
Here only the contribution of the Lorentz
structure $\phi_{\rm B} ({\bf k_{\rm 1}})$ is taken into account, since the
contribution of the second Lorentz structure $\bar{\phi}_{\rm B}$ is
numerically small and has been neglected.
We adopted the
B-meson distribution amplitude widely used in the pQCD approach
\cite{ppnp51-85,prd86-114025,prd87-097501}
\beq
\phi_{\rm B}(x,b)&=& N_{\rm B} x^{\rm 2}(1-x)^{\rm 2}\mathrm{\exp} \left
 [ -\frac{m_{\rm B}^{\rm 2}\; x^{\rm 2}}{2 \omega_{\rm B}^{\rm 2}} -\frac{1}{2} (\omega_{\rm B} b)^{\rm 2}\right],
\label{eq:phib}
\eeq
where the shape parameter $\omega_{\rm B} =0.40$~GeV has
been fixed \cite{ppnp51-85} from the fit to the $B \to \pi$ form
factors derived from lattice QCD and from Light-cone sum rule. In
order to analyze the uncertainties of theoretical predictions
induced by the inputs, we will set $\omega_{\rm B} =0.40 \pm 0.04$~GeV.
The normalization factor $N_{\rm B}$ depends on the values of the shape
parameter $\omega_{\rm B}$ and the decay constant $f_{\rm B}$ and defined
through the normalization relation: $\int_0^1dx\; \phi_{\rm B}(x,b=0)=f_{\rm B}/(2\sqrt{6})$.

For the pseudoscalar $D$ meson and the vector $D^*$ meson, their wave function
can be chosen as \cite{prd78-014018}
\beq
\Phi_{\rm D}(p,x)&=&\frac{i}{\sqrt{6}}\gamma_5 (\psl_{\rm D}+ m_{\rm D} )\phi_{\rm D}(x), \label{eq:wfd} \\
\Phi_{\rm D^*}(p,x) &=& \frac{-i}{\sqrt{6}} \left
 [  \epsl_{\rm L}(\psl_{\rm D^*} +m_{\rm D^*})\phi^L_{\rm D^*}(x)
 + \epsl_{\rm T}(\psl_{\rm D^*} + m_{\rm D^*})\phi^T_{\rm D^*}(x)\right ]
\label{eq:wfdstar}.
 \eeq
 For the distribution amplitudes of $D^{(*)}$ meson, we adopt the one as defined in
Ref.~\cite{prd78-014018}
\beq
\phi_{\rm D^{(*)} }(x)=\frac{ f_{\rm D^{(*)}}}{2\sqrt{6}} 6x(1-x) \left[ 1+C_{D^{(*)} }(1-2x)\right]\exp
\left[-\frac{\omega^{\rm 2} b^{\rm 2} }{2}\right].
 \label{eq:phid}
\eeq
From the heavy quark limit,  we here assume that
$f^{\rm L}_{\rm D^*}=f^{\rm T}_{\rm D^*}=f_{\rm D^*}$,~$\phi^{\rm L}_{\rm D^*}=\phi^{\rm T}_{\rm D^*}=\phi_{\rm D^*}$,
and set $C_{\rm D}=C_{\rm D^*}=0.5,~ \omega=0.1$ GeV as Ref.~\cite{prd78-014018}.

\section*{3 \hspace{0.3cm} Form Factors and Semileptonic Decays}\label{sec:3}

For the semileptonic decays $B \to D l\bar{\nu}_{\rm l}$, the quark level
transitions are $b\to cl\bar\nu_{\rm l}$ decays with the effective
Hamiltonian
\beq \label{eq:hamiltonian}
{\cal H}_{\rm eff}(b\to cl\bar \nu_{\rm l})=\frac{G_{\rm F}}{\sqrt{2}}V_{\rm cb}\; \bar{c}
\gamma_{\mu}(1-\gamma_5)b \cdot \bar l\gamma^{\mu}(1-\gamma_5)\nu_{\rm l},
\eeq
where $G_{\rm F}=1.166 37\times10^{-5} $ GeV$^{-2}$ is the
Fermi-coupling constant.

For $B\to D$ transition,  the form factors $F_{\rm 0,+}(q^{\rm 2})$
can be written in terms of $f_{\rm 1,2}(q^{\rm 2})$ as in Ref.~\cite{prd86-114025}:
\beq \label{eq:f+f0}
F_+(q^{\rm 2})&=&\frac{1}{2}\left[f_{\rm 1}(q^{\rm 2})+f_{\rm 2}(q^{\rm 2})\right], \non
F_0(q^{\rm 2})&=&\frac{1}{2} f_{\rm 1}(q^{\rm 2})\left[1+\frac{q^{\rm 2}}{m_{\rm B}^{\rm 2}-m_{\rm D}^{\rm 2}}\right]
+\frac{1}{2} f_{\rm 2}(q^{\rm 2})\left[1-\frac{q^{\rm 2}}{m_{\rm B}^{\rm 2}-m_{\rm D}^{\rm 2}}\right],
\label{eq:ffdef}
\eeq
with
\beq
f_{\rm 1}(q^{\rm 2})&=&8\pi m^{\rm 2}_{\rm B} C_{\rm F}\int dx_{\rm 1} dx_{\rm 2}\int b_{\rm 1} db_{\rm 1} b_{\rm 2}
db_{\rm 2} \phi_{B}(x_{\rm 1},b_{\rm 1}) \phi_{\rm D}(x_{\rm 2},b_{\rm 2})\non
&& \times \Bigl\{\left[ 2 r\left(1-rx_{\rm 2}\right)
 \right]\cdot h_{\rm 1}(x_{\rm 1},x_{\rm 2},b_{\rm 1},b_{\rm 2})\cdot \alpha_s(t_{\rm 1})\cdot
 \exp\left [-S_{\rm ab}(t_{\rm 1}) \right ] \non
&&+ \left[2 r(2 r_{\rm c}-r)+x_{\rm 1} r \left (-2+2 \eta+\sqrt{\eta^{\rm 2}-1}
-\frac{2 \eta}{\sqrt{\eta^{\rm 2}-1}}+\frac{\eta^{\rm 2}}{\sqrt{\eta^{\rm 2}-1}}
\right ) \right]\non
 && \cdot h_{\rm 2}(x_{\rm 1},x_{\rm 2},b_{\rm 1},b_{\rm 2}) \cdot \alpha_{\rm s} (t_{\rm 2})\cdot
 \exp\left [-S_{\rm ab}(t_{\rm 2}) \right] \Bigr \},
\label{eq:f1q2}
\eeq
\beq
f_{\rm 2}(q^{\rm 2})&=&8\pi m^{\rm 2}_{\rm B} C_{\rm F}\int dx_{\rm 1} dx_{\rm 2}\int b_{\rm 1} db_{\rm 1} b_{\rm 2} db_{\rm 2}
\phi_{\rm B}(x_{\rm 1},b_{\rm 1})\phi_{\rm D}(x_{\rm 2},b_{\rm 2})\non
&& \hspace{-1.2cm}\times \Bigl\{ \left[ 2-4 x_{\rm 2} r(1-\eta) \right]
\cdot h_{\rm 1}(x_{\rm 1},x_{\rm 2},b_{\rm 1},b_{\rm 2})\cdot
\alpha_{\rm s}(t_{\rm 1}) \cdot \exp\left [-S_{\rm ab}(t_{\rm 1}) \right ] \non
&&\hspace{-1.2cm} +  \left[ 4r-2r_{\rm c}-x_{\rm 1}+\frac{x_{\rm 1}}{\sqrt{\eta^{\rm 2}-1}}(2-\eta) \right]
 \cdot h_{\rm 2}(x_{\rm 1},x_{\rm 2},b_{\rm 1},b_{\rm 2}) \cdot \alpha_{\rm s} (t_{\rm 2})\cdot
 \exp\left [-S_{\rm ab}(t_{\rm 2}) \right]
 \Bigr \}, \quad
\label{eq:f2q2}
 \eeq
where $C_{\rm F}=4/3$ is a color factor, $r_{\rm c}=m_{\rm c}/m_{\rm B}$ with $m_{\rm c}$ is the mass
of $c$-quark. The hard functions $h_{\rm 1,2}(x_{\rm i},b_{\rm i})$ come
form the Fourier transform and can be written as ~\cite{prd65-014007,prd63-074009}
\beq
h_{\rm 1}(x_{\rm 1},x_{\rm 2},b_{\rm 1},b_{\rm 2})&=&
K_{\rm 0}(\beta_{\rm 1} b_{\rm 1}) \left \{\theta(b_{\rm 1}-b_{\rm 2})I_{\rm 0}(\alpha_{\rm 1} b_{\rm 2})
K_{\rm 0}(\alpha_{\rm 1} b_{\rm 1})\right. \non && \left. +\theta(b_{\rm 2}-b_{\rm 1})I_{\rm 0}(\alpha_{\rm 1}
b_{\rm 1})K_{\rm 0}(\alpha_{\rm 1} b_{\rm 2}) \right \}\cdot S_{\rm t}(x_{\rm 2}),\non
h_{\rm 2}(x_{\rm 1},x_{\rm 2},b_{\rm 1},b_{\rm 2})&=&K_{\rm 0}(\beta_{\rm 2} b_{\rm 1}) \left \{
\theta(b_{\rm 1}-b_{\rm 2})I_{\rm 0}(\alpha_{\rm 2} b_{\rm 2})K_{\rm 0}(\alpha_{\rm 2} b_{\rm 1})\right.\non &&
\left. +\theta(b_{\rm 2}-b_{\rm 1})I_{\rm 0}(\alpha_{\rm 2} b_{\rm 1})K_{\rm 0}(\alpha_{\rm 2} b_{\rm 2}) \right \}
\cdot S_{\rm t}(x_{\rm 2}),
\label{eq:h1h2}
\eeq
where $K_{\rm 0}$ and $I_{\rm 0}$ are modified Bessel functions, while the parameters
\beq
\alpha_{\rm 1} = m_{B}\sqrt{x_{\rm 2}r \eta^+},\quad \alpha_{\rm 2}=m_{\rm B}\sqrt{x_{\rm 1} r \eta^+ -
r^{\rm 2}+r_{\rm c}^{\rm 2}},\quad \beta_{\rm 1} = \beta_{\rm 2}=m_{\rm B}\sqrt{x_{\rm 1}x_{\rm 2} r \eta^+},
\eeq
with $r=m_{D^{(*)}}/m_{\rm B}$. The threshold resummation factor
$S_{\rm t}(x_{\rm i})$ is adopted from \cite{prd65-014007}, and the Sudakov factor
$S_{\rm ab}(t)=S_{\rm B}(t)+S_{\rm M}(t)$ can be found in Refs.~\cite{prd65-014007,prd63-074009}.

With the form factors $F_{\rm 0,+}(q^{\rm 2})$,
the differential decay widths of the semileptonic decays
$B \to D l\bar{\nu}_{\rm l}$  can be written as \cite{prd79-014013}
\beq
\frac{d\Gamma(B \to D l\bar{\nu}_{\rm l})}{dq^{\rm 2}}&=&\frac{G_F^{\rm 2}|V_{\rm cb}|^{\rm 2}}{192 \pi^3  m_{\rm B}^3}
\left ( 1-\frac{m_{\rm l}^{\rm 2}}{q^{\rm 2}} \right)^{\rm 2}\frac{\lambda^{1/2}(q^{\rm 2})}{2q^{\rm 2}}
\non
&& \hspace{-1cm} \cdot
 \Bigl \{  3 m_{\rm l}^{\rm 2}\left (m_{\rm B}^{\rm 2}-m_{\rm D}^{\rm 2}
\right )^{\rm 2} |F_{\rm 0}(q^{\rm 2})|^{\rm 2}
+ \left (m_{\rm l}^{\rm 2}+2q^{\rm 2} \right )\lambda(q^{\rm 2})|F_+(q^{\rm 2})|^{\rm 2} \Bigr \}, \label{eq:dg1}
\eeq
where $m_{\rm l}$ is the mass of the charged leptons, and
$\lambda(q^{\rm 2}) = (m_{\rm B}^{\rm 2}+m_{\rm D}^{\rm 2}-q^{\rm 2})^{\rm 2} - 4 m_{\rm B}^{\rm 2} m_{\rm D}^{\rm 2}$ is the phase space factor.

For $B \to D^*$ transitions, the relevant form factors are
$V(q^{\rm 2})$ and $A_{\rm 0,1,2}(q^{\rm 2})$ \cite{prd65-014007}.
By employing the pQCD approach, we calculate and find the
expressions for these form factors:
\beq
V(q^{\rm 2})&=&8\pi m^{\rm 2}_{\rm B} C_{\rm F}\int dx_{\rm 1} dx_{\rm 2}\int b_{\rm 1} db_{\rm 1} b_{\rm 2} db_{\rm 2}
\phi_{\rm B}(x_{\rm 1},b_{\rm 1})\phi^{\rm T}_{\rm D^*}(x_{\rm 2},b_{\rm 2}) \cdot (1+r)\non && \times \Bigl
\{\left[1-rx_{\rm 2}\right] \cdot h_{\rm 1}(x_{\rm 1},x_{\rm 2},b_{\rm 1},b_{\rm 2})\cdot \alpha_{\rm s}(t_{\rm 1})
\cdot \exp\left [-S_{\rm ab}(t_{\rm 1}) \right ] \non && + \left[
r+\frac{x_{\rm 1}}{2\sqrt{\eta^{\rm 2}-1}}\right] \cdot h_{\rm 2}(x_{\rm 1},x_{\rm 2},b_{\rm 1},b_{\rm 2})
\cdot \alpha_{\rm s} (t_{\rm 2})\cdot \exp\left [-S_{\rm ab}(t_{\rm 2})\right] \Bigr \},
\label{eq:Vqq}
\eeq
\beq
A_{\rm 0}(q^{\rm 2})&=&8\pi m^{\rm 2}_{\rm B} C_{\rm F}\int dx_{\rm 1} dx_{\rm 2}\int b_{\rm 1} db_{\rm 1} b_{\rm 2} db_{\rm 2}
\phi_{\rm B}(x_{\rm 1},b_{\rm 1})\phi^{\rm L}_{\rm D^*}(x_{\rm 2},b_{\rm 2})\non && \times \Bigl \{ \left[
1+r -rx_{\rm 2}(2+r-2\eta)\right]\cdot h_{\rm 1}(x_{\rm 1},x_{\rm 2},b_{\rm 1},b_{\rm 2})\cdot
\alpha_{\rm s}(t_{\rm 1}) \cdot \exp\left [-S_{\rm ab}(t_{\rm 1}) \right ] \non && + \left
[r^{\rm 2}+r_{\rm c}+\frac{x_{\rm 1}}{2}+\frac{\eta x_{\rm 1}}{
2\sqrt{\eta^{\rm 2}-1}}+\frac{rx_{\rm 1}}{2\sqrt{\eta^{\rm 2}-1}} \left
(1-2\eta(\eta+\sqrt{\eta^{\rm 2}-1}) \right )\right ] \non && \cdot
h_{\rm 2}(x_{\rm 1},x_{\rm 2},b_{\rm 1},b_{\rm 2}) \cdot \alpha_{\rm s} (t_{\rm 2})\cdot \exp\left
[-S_{\rm ab}(t_{\rm 2}) \right] \Bigr \},
 \label{eq:A0qq}
\eeq
\beq
A_{\rm 1}(q^{\rm 2})&=&8\pi m^{\rm 2}_{\rm B} C_{\rm F}\int dx_{\rm 1} dx_{\rm 2}\int b_{\rm 1} db_{\rm 1} b_{\rm 2} db_{\rm 2}
\phi_{\rm B}(x_{\rm 1},b_{\rm 1})\phi^{\rm T}_{\rm D^*}(x_{\rm 2},b_{\rm 2})\cdot \frac{r}{1+r}\non &&
\times  \Bigl \{2 [ 1+\eta-2 r x_{\rm 2}+r\eta x_{\rm 2} ]\cdot
h_{\rm 1}(x_{\rm 1},x_{\rm 2},b_{\rm 1},b_{\rm 2})\cdot \alpha_{\rm s}(t_{\rm 1})\cdot \exp[-S_{\rm ab}(t_{\rm 1})]\non
& & +  \left[ 2r_{\rm c}+2 \eta r-x_{\rm 1}\right]\cdot h_{\rm 2}(x_{\rm 1},x_{\rm 2},b_{\rm 1},b_{\rm 2})
\cdot \alpha_{\rm s} (t_{\rm 2})\cdot \exp[-S_{\rm ab}(t_{\rm 2})] \Bigr \},
\label{eq:A1qq}
\eeq
\beq
A_{\rm 2}(q^{\rm 2})&=&\frac{(1+r)^{\rm 2}(\eta-r)}{2r(\eta^{\rm 2}-1)}\cdot A_{\rm 1}(q^{\rm 2})- 8\pi
m^{\rm 2}_{\rm B} C_{\rm F}\int dx_{\rm 1} dx_{\rm 2}\int b_{\rm 1} db_{\rm 1} b
_{\rm 2} db_{\rm 2}\phi_{\rm B}(x_{\rm 1},b_{\rm 1})
\non & & \cdot \phi^{\rm L}_{\rm D^*}(x_{\rm 2},b_{\rm 2}) \cdot \frac{1+r}{\eta^{\rm 2}-1}
\times \Bigl \{  \left[(1+\eta)(1-r)
-rx_{\rm 2}(1-2r+\eta(2+r-2\eta))\right]\non && \cdot
h_{\rm 1}(x_{\rm 1},x_{\rm 2},b_{\rm 1},b_{\rm 2})\cdot \alpha_{\rm s}(t_{\rm 1}) \cdot \exp\left
[-S_{\rm ab}(t_{\rm 1}) \right ] \non &+& \left[r+r_{\rm c}(\eta-r)-\eta r^{\rm 2}+r x_{\rm 1}
\eta^{\rm 2}-\frac{x_{\rm 1}}{2}(\eta+r)+x_{\rm 1} \left (\eta r -\frac{1}{2} \right)
\sqrt{\eta^{\rm 2}-1}\right]\non &\cdot& h_{\rm 2}(x_{\rm 1},x_{\rm 2},b_{\rm 1},b_{\rm 2}) \cdot
\alpha_{\rm s} (t_{\rm 2})\cdot \exp\left [-S_{\rm ab}(t_{\rm 2}) \right] \Bigr \},
\label{eq:A2qq}
\eeq
where $r=m_{\rm D^*}/m_{\rm B}$, while $C_{\rm F}$ and $r_{\rm c}$ is the same as in
Eqs.~(\ref{eq:f1q2},\ref{eq:f2q2}).

For $B \to D^* l\bar{\nu}_{\rm l}$ decays,  the differential decay
widths can be written as \cite{prd79-054012}
\beq
\frac{d\Gamma_{\rm L}(B \to D^*l\bar{\nu}_{\rm l})}{dq^{\rm 2}}&=&
\frac{G_{\rm F}^{\rm 2}|V_{\rm cb}|^{\rm 2}}{192 \pi^3  m_{\rm B}^3} \left ( 1-\frac{m_{\rm l}^{\rm 2}}{q^{\rm 2}}\right )^{\rm 2}
\frac{\lambda^{1/2}(q^{\rm 2})}{2q^{\rm 2}}\cdot \Bigg\{3m^{\rm 2}_{\rm l}\lambda(q^{\rm 2})A^{\rm 2}_{\rm 0}(q^{\rm 2})\non
&& \hspace{-2cm}
+\frac{m^{\rm 2}_{\rm l}+2q^{\rm 2}}{4m^{\rm 2}}\cdot \left [(m^{\rm 2}_{\rm B}-m^{\rm 2}-q^{\rm 2})(m_{\rm B}+m)A_{\rm 1}(q^{\rm 2})
 -\frac{\lambda(q^{\rm 2})}{m_{\rm B}+m}A_{\rm 2}(q^{\rm 2}) \right ]^{\rm 2} \Bigg\},
\label{eq:dfds1}
\eeq
\beq
\frac{d\Gamma_\pm(B \to D^* l\bar\nu_{\rm l})}{dq^{\rm 2}}&=&
\frac{G_F^{\rm 2}|V_{\rm cb}|^{\rm 2}}{192 \pi^3
 m_{\rm B}^3}\left ( 1-\frac{m_{\rm l}^{\rm 2}}{q^{\rm 2}}\right )^{\rm 2} \frac{\lambda^{3/2}(q^{\rm 2})}{2}\non
& & \cdot \left \{ (m^{\rm 2}_{\rm l}+2q^{\rm 2})\left[\frac{V(q^{\rm 2})}{m_{\rm B}+m}\mp
\frac{(m_{\rm B}+m)A_{\rm 1}(q^{\rm 2})}{\sqrt{\lambda(q^{\rm 2})}}\right]^{\rm 2}\right\},
\label{eq:dfds2}
\eeq
where $m=m_{\rm D^*}$, and $\lambda(q^{\rm 2}) = (m_{\rm B}^{\rm 2}+m_{\rm D^*}^{\rm 2}-q^{\rm 2})^{\rm 2} - 4 m_{\rm B}^{\rm 2} m_{\rm D^*}^{\rm 2}$
is the phase space factor. The total differential decay widths is defined as
\beq
\frac{d\Gamma}{dq^{\rm 2}}=\frac{d\Gamma_{\rm L}}{dq^{\rm 2}} +
\frac{d\Gamma_+}{dq^{\rm 2}}    +\frac{d\Gamma_-}{dq^{\rm 2}}\; .
\label{eq:dfdst}
\eeq

\section*{4 \hspace{0.3cm} Numerical Results and Discussions} \label{sec:4}

In the numerical calculations we use the following input parameters
(here masses and decay constants in units of GeV)\cite{pdg2012,prd86-114025,hfag2012}:
\beq
 m_{\rm D^0}&=&1.865, \quad m_{\rm D^+}=1.870, \quad
m_{\rm D^{0*}}=2.007,\quad m_{\rm D^{*+}}=2.010,\quad m_{\rm B}=5.28,\non
m_{\tau}&=& 1.777,\quad m_{c}=1.35 \pm 0.03, \quad f_{\rm B}=0.21\pm0.02, \quad f_{\rm D}=0.223,  \non
|V_{\rm cb}|&=&(39.54\pm 0.89)\times 10^{-3}, \quad \Lambda^{\rm (f=4)}_{\overline{\rm MS}} = 0.287,\non
\tau_{\rm B^\pm}&=&1.641\; {\rm ps},\quad \tau_{\rm B^0}=1.519\; {\rm ps}, \quad
f_{\rm D^*}= f_{\rm D}\sqrt{m_{\rm D}/m_{\rm D^*}}.
\label{eq:inputs}
\eeq

\subsection*{4.1 \hspace{0.3cm}Form factors in the pQCD factorization approach }

For the considered semileptonic decays, the differential decay rates
strongly depend on the value and the shape of the relevant
form factors $F_{\rm 0,+}(q^{\rm 2})$, $V(q^{\rm 2})$ and $A_{\rm 0,1,2}(q^{\rm 2})$.
Besides the two well-known traditional methods of evaluating the form factors,
the QCD sum rule for the low $q^{\rm 2}$ region and the Lattice QCD for the high
$q^{\rm 2}$ region of $q^{\rm 2}\approx q_{\rm max}^{\rm 2}$,
one can also calculate the form factors perturbatively in the
low $q^{\rm 2}$ region by employing the pQCD factorization approach
\cite{li1995,prd67-054028,prd85-094003,prd86-114025,R27,prd78-014018,
prd65-014007,prd63-074009,pqcd-ff,R37}.

In Refs.~\cite{li1995,prd67-054028,prd78-014018},
the authors examined the applicability of the pQCD approach to
$B \to (D, D^*)$ transitions, and have shown that the pQCD
approach with the inclusion of the Sudakov effects is  applicable to the
semileptonic decays $B \to D^{(*)} l\bar{\nu}_{\rm l}$ in the lower
$q^{\rm 2}$ region (i.e. the $D$ or $D^*$ meson recoils fast).
Since the pQCD predictions for the considered form factors are reliable
only for small values of $q^{\rm 2}$, we will calculate explicitly
the values of the form factors $F_{\rm 0,+}(q^{\rm 2})$, $V(q^{\rm 2})$
and $A_{\rm 0,1,2}(q^{\rm 2})$ in the lower range of $ m_{\rm l}^{\rm 2} \leq q^{\rm 2} \leq m_{\rm \tau}^{\rm 2}$
with $l=(e,\mu)$ by  using the expressions as given in
Eqs.(\ref{eq:f1q2},\ref{eq:f2q2},\ref{eq:Vqq}-\ref{eq:A2qq}) and the
definitions in Eq.~(\ref{eq:f+f0}).

In Table \ref{tab:ffab}, we list the pQCD predictions for all relevant
form factors for $B \to D^{(*)}$ transitions at the points $q^{\rm 2}=0$ and
$q^{\rm 2}=m_{\rm \tau}^{\rm 2}$, respectively.
The total error of the pQCD predictions is the combination
of the major errors from the uncertainty of $\omega_{\rm B}=0.40\pm 0.04$ GeV,
$f_{\rm B}=0.21\pm 0.02$ GeV and $m_{\rm c}=1.35 \pm 0.03$ GeV.

\begin{table}[thb]
\begin{center}
\caption{The pQCD predictions for the form factors $F_{\rm 0,+}, V$ and
$A_{\rm 0,1,2}$ at $q^{\rm 2}=0, m_{\rm \tau}^{\rm 2}$, and the parametrization constants ``a" and ``b" for
$B \to D$ and $B \to D^*$ transitions.}
\label{tab:ffab}
\vspace{0.2cm}
\begin{tabular}{l llll} \hline \hline
& $F(0)$ & $F(m_{\rm \tau}^{\rm 2})$ & $a$ & $b$ \\ \hline \hline
$F_{\rm 0}^{\rm B\to D}$ & $0.52^{+0.12}_{-0.10}$& $0.64^{+0.14}_{-0.12}$
&$1.71^{+0.05}_{-0.07}$& $0.52^{+0.13}_{-0.11}$ \\ \hline
$F_+^{\rm B\to D}$ & $0.52^{+0.12}_{-0.10}$ &$0.70^{+0.16}_{-0.14}$&
$2.44^{+0.04}_{-0.05}$& $1.49^{+0.09}_{-0.09}$ \\ \hline \hline
$V^{\rm B\to D^*}$ & $0.59^{+0.12}_{-0.11}$ &$0.79^{+0.15}_{-0.14}$&
$2.41^{+0.13}_{-0.17}$& $1.76^{+0.14}_{-0.01}$ \\
\hline $A_{\rm 0}^{\rm B\to D^*}$ & $0.46^{+0.10}_{-0.08}$&$0.62^{+0.12}_{-0.11}$ &$2.44^{+0.10}_{-0.14}$&$1.98^{+0.05}_{-0.10}$ \\
\hline $A_{\rm 1}^{\rm B\to D^*}$ & $0.48^{+0.10}_{-0.09}$&$0.58^{+0.11}_{-0.10}$&
$1.61^{+0.13}_{-0.15}$&$0.75^{+0.15}_{-0.09}$ \\
\hline $A_{\rm 2}^{\rm B\to D^*}$ & $0.51^{+0.11}_{-0.09}$&$0.66^{+0.13}_{-0.12}$&
$2.29^{+0.13}_{-0.15}$&$1.89^{+0.11}_{-0.09}$ \\
\hline \hline
\end{tabular}
\end{center} \end{table}

In the lower $q^{\rm 2}$ region of $0\leq q^{\rm 2} \leq m_{\rm \tau}^{\rm 2}$, we firstly calculate the form
factors $F_{\rm 0,+}(q^{\rm 2})$ for $B \to D$ transition at the sixteen points by employing the
pQCD approach respectively.
Secondly we make an extrapolation for the form factors $F_{\rm 0,+}(q^{\rm 2})$
from the lower $q^{\rm 2}$ region to the larger $q^{\rm 2}$
region $m_{\rm \tau}^{\rm 2} < q^{\rm 2} \leq q_{\rm max}^{\rm 2}=(m_{\rm B}-m_{\rm D})^{\rm 2}$ by using the pole model
parametrization ~\cite{cheng2004,prd79-054012}
\beq
F_{\rm 0,+}(q^{\rm 2})=\frac{F_{\rm 0,+}(0)}{1-a (q^{\rm 2}/m_{\rm B}^{\rm 2}) + b\left( q^{\rm 2}/m_{\rm B}^{\rm 2}\right)^{\rm 2}}.
\label{eq:fq2}
\eeq
The parameters $a$ and $b$ in above equation are determined
by the fitting to the pQCD predicted values obtained at the sixteen points in
the lower $q^{\rm 2}$ region, and have been given in Table \ref{tab:ffab}.
For the form factors $V(q^{\rm 2})$ and $A_{\rm 0,1,2}(q^{\rm 2})$ we
show the numerical  results also in Table \ref{tab:ffab}.

\subsection*{4.2 \hspace{0.3cm} Differential decay widths and branching ratios }

\begin{figure}[thb]
\begin{center}
\includegraphics[scale=0.2]{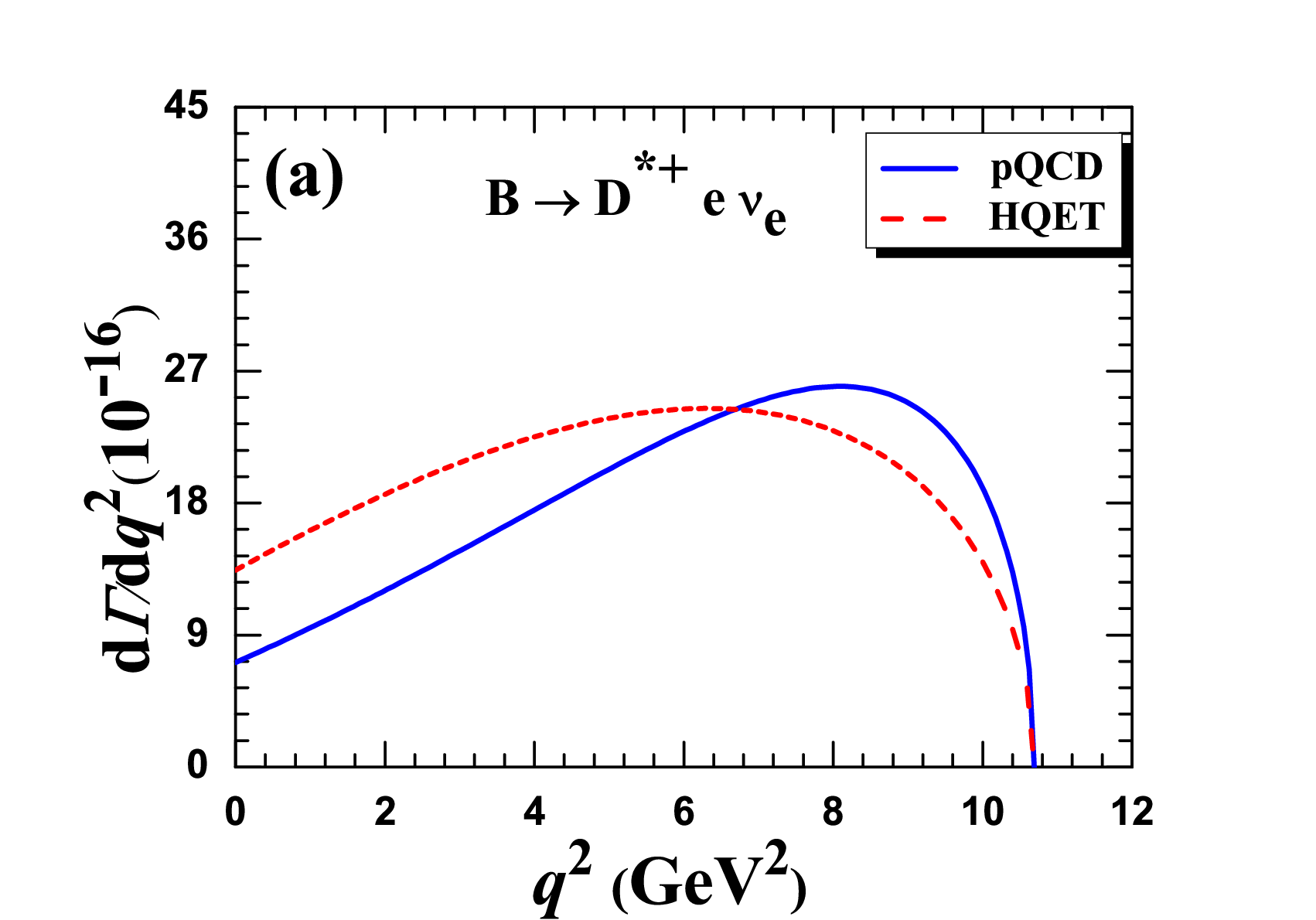} \hspace{-0.6cm}
\includegraphics[scale=0.2]{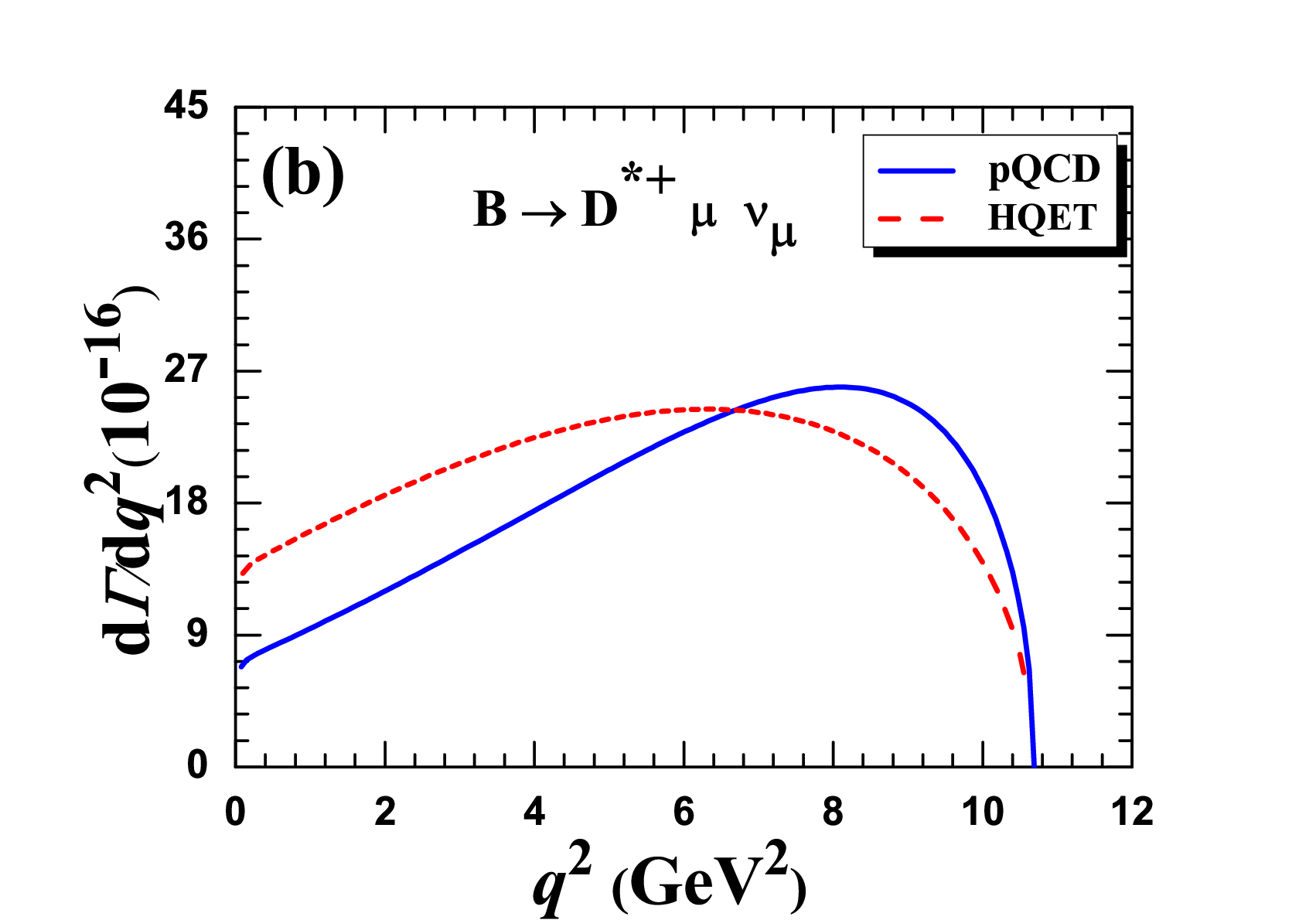} \hspace{-0.6cm}
\includegraphics[scale=0.2]{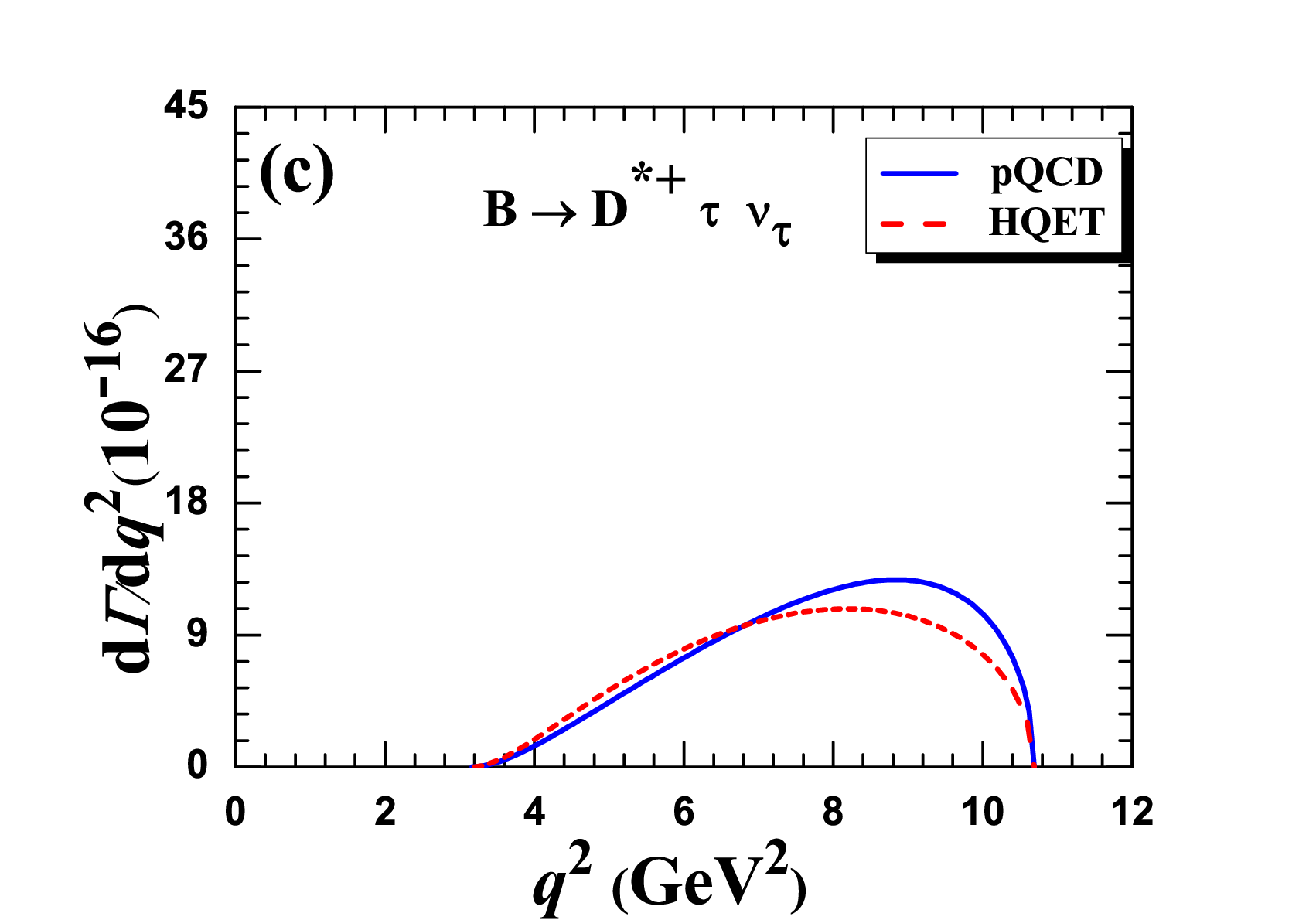}
\caption{The $q^{\rm 2}$-dependence of $d\Gamma/dq^{\rm 2}$ for
$B \to D^{*+} l^-\bar{\nu}_{\rm l} $ with $l=e,\mu,\tau$ in the pQCD
approach (the solid curves),
or in the HQET (the short-dashed curves). Here $q^{\rm 2}_{\rm max}$ = 10.69 GeV$^{\rm 2}$.}
\label{fig:fig2}
\end{center}
\end{figure}

In Fig.~\ref{fig:fig2}, we show the pQCD approach for $q^{\rm 2}$-dependence of the
theoretical predictions for $d\Gamma/dq^{\rm 2}$ for $B\to D^{*+} l^- \bar{\nu}_{\rm l}$
decays (the solid curves ) or the traditional HQET method
\cite{HQET,R40,R41,R42,grozin-2004,prd85-094025,jhep2013-01054,Caprini:1997mu}.
From these figures one can see that:
\begin{itemize}
\item[1)]
For $B\to D^{*+} l^- \bar{\nu}_{\rm l}$ ($l=e,\mu$) decays, the pQCD predictions for
$d\Gamma/dq^{\rm 2}$ become larger  than the HQET ones in the region of
$ q^{\rm 2} > 6.5$ GeV$^{\rm 2}$, and then approach zero at the end
point $q^{\rm 2}_{\rm max}=(m_{\rm B}-m_{D^{*+}})^{\rm 2}=10.69$ GeV$^{\rm 2}$.

\item[2)]
For $B\to D^{(*)} \tau \bar{\nu}_{\rm \tau}$ decays, the difference between
the pQCD and HQET predictions for $d\Gamma/dq^{\rm 2}$ are small in the whole range
of $q^{\rm 2}$.
\end{itemize}
For other decays we have very similar pQCD predictions  for $q^{\rm 2}$-dependence
of the differential decay widths.

From the differential decay rates as given
in Eqs.(\ref{eq:dg1},\ref{eq:dfds1}-\ref{eq:dfdst}),
it is easy to calculate the branching ratios for the considered decays
by the integrations over $q^{\rm 2}$. We then find the pQCD predictions
for the branching ratios of the eight $B \to D^{(*)} l^-
\bar{\nu}_{\rm l}$ decays as listed in Table \ref{tab:br31}, where the
individual theoretical errors have been added in quadrature. In
order to check the relative size of the theoretical errors, we here
present the pQCD predictions for ${\cal B}(B^- \to D^0 \tau^-\bar{\nu}_{\rm \tau})$
and ${\cal B}(B^- \to D^{*0} \tau^-\bar{\nu}_{\rm \tau})$  with the individual errors:
\beq
 {\cal B}(B^- \to D^0 \tau^- \bar{\nu}_{\rm \tau}) &=&
\left [ 0.95^{+0.31}_{-0.25}(\omega_{\rm B})^{+0.19}_{-0.17}
(f_{\rm B})\pm 0.04(V_{\rm cb}) \pm 0.01 (m_{\rm c}) \right ]\%, \label{eq:br11}\\
{\cal B}(B^-\to D^{*0} \tau^- \bar{\nu}_{\rm \tau}) &=&
\left [ 1.47^{+0.29}_{-0.27}(\omega_{\rm B})^{+0.29}_{-0.27}
(f_{\rm B}) \pm 0.07(V_{\rm cb}) \pm 0.10(m_{\rm c}) \right]\%,
\label{eq:br12}
 \eeq
where the four major theoretical errors come from the uncertainties
of the input parameters $\omega_{\rm B}=0.40\pm
0.04$ GeV, $f_{\rm B}=0.21\pm 0.02$ GeV, $|V_{\rm cb}|=(39.54\pm 0.89)
\times 10^{-3}$ and $m_{\rm c}=1.35\pm 0.03$ GeV.

In Table \ref{tab:br31}, the pQCD predictions for the branching ratios of the eight decay
modes are listed in column two. For the case of $l=(e,\mu)$, we list the
averaged results.
In column three, we show the HQET predictions obtained by direct calculations
using the formulaes as given in Refs.~\cite{prd85-094025,jhep2013-01054}
The HQET predictions as given in Ref.~\cite{prd85-094025} are listed
in column four.
The measured values as reported by BaBar \cite{prl109-101802} or quoted
from PDG-2012 \cite{pdg2012} are also listed in last two columns as an comparison.
One can see from the numerical results in Table \ref{tab:br31} that
\begin{itemize}
\item[1)]
The pQCD and HQET predictions for the branching ratios
in fact agree with each other within one standard deviation, but the central values
of the pQCD predictions for  the branching ratios  of the four
$B \to D^{(*)} \tau \bar{\nu}_{\rm \tau}$ decays
are a little larger than the HQET ones and show a better agreement with the
measured values.

\item[2)]
Of course, the theoretical errors of the pQCD predictions are still large,
say $\sim 35\%$.
It is therefore necessary to define the ratios $R(X)$ among the branching ratios
of the individual decays, since the theoretical errors are greatly
canceled in these ratios.

\end{itemize}

\begin{table}[thb]
\begin{center}
\caption{ The theoretical predictions for  ${\cal B}(B\to D^{(*)} l^- \bar{\nu}_{\rm l})$.
The world averages from PDG 2012 \cite{pdg2012} and the measured
values ~\cite{prl109-101802,prd79-012002,prd77-032002} are also
listed in last two columns. } \label{tab:br31}
\vspace{0.2cm}
\begin{tabular}{l ll l l l} \hline \hline
~~ Channels~~ & pQCD(\%) \hspace{0.3cm} & HQET(\%)\hspace{0.3cm} &  HQET(\%)[5]\hspace{0.3cm}
& PDG(\%)[34]\hspace{0.3cm} & BaBar(\%)\hspace{0.3cm}   \\ \hline
$\bar{B}^0 \to D^+ \tau^- \bar{\nu}_{\rm \tau}$ & $0.87^{+0.34}_{-0.28}$ & $0.63 \pm 0.06$
&$0.64 \pm 0.05 $ &  $1.1 \pm 0.4 $ & $1.01 \pm 0.22$\\
$\bar{B}^0 \to D^+ l^- \bar{\nu}_{\rm l}$ & $2.03^{+0.92}_{-0.70}$ &$2.13 ^{+0.19}_{-0.18}$
& $-$ & $2.18 \pm 0.12 $ & $2.15 \pm 0.08$\\
$B^- \to D^0 \tau^- \bar{\nu}_{\rm \tau}$ & $0.95^{+0.37}_{-0.31}$ &$0.69 \pm 0.06$
& $0.66 \pm 0.05 $ & $0.77 \pm 0.25 $  & $0.99 \pm 0.23$\\
$B^- \to D^0 l^- \bar{\nu}_{\rm l}$ & $2.19^{+0.99}_{-0.76}$ &$2.30 \pm 0.20$
& $-$ & $2.26 \pm 0.11 $ & $2.34\pm 0.14$\\ \hline
$\bar{B}^0 \to D^{*+}\tau^-\bar{\nu}_{\rm \tau}$ & $1.36^{+0.38}_{-0.37}$ &$1.25\pm 0.04$
& $1.29 \pm 0.06 $ & $1.5 \pm 0.5$&$1.74\pm 0.23$\\
$\bar{B}^0\to D^{*+} l^- \bar{\nu}_{\rm l}$ & $4.52^{+1.44}_{-1.31}$ &$4.94\pm 0.15$
& $-$& $4.95 \pm 0.11$&$4.69\pm 0.34$\\
$B^-\to D^{*0} \tau^- \bar{\nu}_{\rm \tau}$ & $1.47^{+0.43}_{-0.40}$ &$1.35 \pm 0.04$
& $1.43 \pm 0.05 $& $2.04 \pm 0.30$&$1.71\pm 0.21$\\
$B^- \to D^{*0} l^- \bar{\nu}_{\rm l}$ & $4.87^{+1.60}_{-1.41}$ &$5.35 \pm 0.16$
& $-$& $5.70 \pm 0.19$&$5.40\pm 0.22$\\
\hline \hline
\end{tabular}\end{center} \end{table}

\subsection*{4.3 \hspace{0.3cm} The ratios of the branching ratios}

Since the most hadronic and SM parameter uncertainties are greatly canceled in the
ratios of the corresponding branching ratios, we firstly define the six R(X)-ratios
in the same way as in Ref.~\cite{prl109-101802} and compare our pQCD predictions with
other theoretical predictions or the measured values.
For the two isospin-constrained ratios $R(D)$ and $R(D^*)$, for example, we find
numerically
\beq
R(D)&=& 0.430^{+0.015}_{-0.022}(\omega_{\rm B}) \pm 0.014 (m_{\rm c}), \label{eq:rdt1}\\
R(D^*) &=& 0.301^{+0.012}_{-0.013}(\omega_{\rm B})^{+0.005}_{-0.003}(m_{\rm c}),
\label{eq:rdt2}
 \eeq
where the major theoretical errors come from
the uncertainties of $\omega_{\rm B}=0.40\pm 0.04$ GeV and $m_{\rm c}=1.35\pm
0.03$ GeV. The theoretical errors from the uncertainties of $f_{\rm B}$ and $|V_{\rm cb}|$
are canceled completely in the ratios of the branching ratios. It is
easy to see that the theoretical errors of the pQCD predictions for
R(X)-ratios are reduced significantly to about $5\%$.

\begin{table}[thb]
\begin{center}
\caption{ The theoretical predictions for the six R-ratios obtained by employing the
pQCD approach  or other theoretical methods, and the measured values \cite{prl109-101802}.}
\label{tab:ratios}\vspace{0.2cm}
\begin{tabular}{l ll lllll} \hline \hline
Ratio &pQCD & HQET &HQET [5]& SM [7,8]\hspace{0.3cm}
&SM [14]\hspace{0.3cm}& SM [19]\hspace{0.3cm} & BaBar [4] \\ \hline
$R(D^0)$   &$0.433^{+0.017}_{-0.027}$&$0.297^{+0.017}_{-0.016}$&$-$&$-$&$-$&$-$&$0.429\pm 0.097$ \\
$R(D^+)$   &$0.428^{+0.023}_{-0.033}$&$0.297\pm 0.017$&$-$&$-$&$-$&$-$&$0.469\pm 0.099$ \\
$R(D^{*0})$&$0.302^{+0.012}_{-0.014}$&$0.253\pm 0.004$&$-$&$-$&$-$&$-$&$0.322\pm 0.039$ \\
$R(D^{*+})$&$0.301^{+0.012}_{-0.015}$&$0.252\pm 0.004$&$-$&$-$&$-$&$-$&$0.355\pm 0.044$ \\ \hline
${\cal R}(D)$  & $0.430^{+0.021}_{-0.026}$ & $0.297\pm 0.017$& $0.296\pm 0.016$&$0.316$&$0.315$ & $0.31$&$0.440 \pm 0.072$\\
${\cal R}(D^*)$& $0.301\pm 0.013         $ & $0.252\pm 0.004$& $0.252\pm 0.003$&$ -$&$0.260$&$-$&$0.332 \pm 0.030 $\\
\hline\hline
\end{tabular}
\end{center} \end{table}

In Table \ref{tab:ratios}, we list our pQCD  predictions for all six R(X)-ratios
in column two. As comparisons, we also show the HQET predictions obtained in this work or
those as given in  Refs.~\cite{prd85-094025}, other SM predictions
as  presented in Refs.~\cite{prl109-071802,prd85-114502,mpla27-1250183,Kosnik-1301},
and the measured values as reported by BaBar Collaboration \cite{prl109-101802}.
From the numerical results as listed in Table II and III we find the following points:
\begin{itemize}
\item[1)]
Due to the strong cancelation of the theoretical errors in the ratios of the
corresponding branching ratios, the error of the pQCD predictions for all six R(X)-ratios
are $\sim 5\%$ only,  similar in size with
the HQET ones (in this work or in Ref.~\cite{prd85-094025}) and  other SM
predictions \cite{prl109-071802,prd85-114502,mpla27-1250183,Kosnik-1301}.

\item[2)]
The SM predictions as given in Refs.~\cite{prl109-071802,prd85-114502,mpla27-1250183,Kosnik-1301}
are consistent with each other within their errors.
One can see that, however, there still exist a clear discrepancy between
these theoretical predictions for $R(D^{(*)})$ and the BaBar's measurements
\cite{prl109-101802},
although the gap become a little bit smaller than that in Ref.~\cite{prd85-094025}.

\item[3)]
For $R(D)$ and $R(D^*)$, the pQCD predictions
agree very well with the data, the BaBar's anomaly
of $R(D^{(*)})$ are therefore explained successfully in the framework of the SM
by employing the pQCD factorization approach.

\item[4)]
Besides $R(D^{(*)})$, the pQCD predictions for the central values of other four
ratios $R(D^0)$, $R(D^+)$, $R(D^{*0})$ and $R(D^{*+})$ also
agree very well with the corresponding BaBar measurements.

\end{itemize}

Analogous to above $R(D)$ and $R(D^*)$ ratios, we can also define the new
isospin-constrained ratios $R_{\rm D}^l$ and $R_{\rm D}^\tau$ in the form of
\beq
R_{\rm D}^{\rm l} &\equiv& \frac{ {\cal B}(B \to D^+l^- \bar{\nu}_{\rm l} ) + {\cal B}(B \to D^0 l^- \bar{\nu}_{\rm l})}{ {\cal B}(B
\to D^{*+} l^- \bar{\nu}_{\rm l} )+ {\cal B}(B \to D^{*0} l^- \bar{\nu}_{\rm l})} , \label{eq:rdt5} \\
R_{\rm D}^{\rm \tau} &\equiv & \frac{ {\cal B}(B \to D^+ \tau^- \bar{\nu}_{\rm \tau} )+ {\cal B}(B \to D^0 \tau^- \bar{\nu}_{\rm \tau})}{
{\cal B}(B \to D^{*+} \tau^- \bar{\nu}_{\rm \tau} )+ {\cal B}(B \to D^{*0} \tau^- \bar{\nu}_{\rm \tau})} .
\label{eq:rdt6}
\eeq

In the ratio $R(D)$ ( $R(D^*)$), the involved decays have the same final state meson
$D$ ( $D^*$), but different leptons.
The value of the ratio $R(D^{(*)})$ dominantly depend on the mass difference
between large $m_{\rm \tau}$ and tiny $m_{\rm l}$ with $l=(e,\mu)$.
For the two new ratios $R_{\rm D}^{\rm l}$ and $R_{\rm D}^{\rm \tau}$, however,
the relevant decays appeared in one ratio have the same final state leptons
but different final state mesons: $D$ in the numerator and $D^*$ in the denominator.
The new ratio $R_{\rm D}^{\rm l}$ and $R_{\rm D}^{\rm \tau}$ will measure the effects induced by
the variations of the form factors for $B \to D$ and $B \to D^*$
transition, respectively. In other words, the new ratios $R_{\rm D}^l$ and $R_{\rm D}^\tau$
may be more sensitive to the QCD dynamics which controls the $B \to D^{(*)}$ transitions
than the ``old" ratios $R(D)$ and $R(D^*)$. We therefore suggest the experimental
measurements for the new ratios $R_{\rm D}^{\rm l}$ and $R_{\rm D}^{\rm \tau}$ as soon as possible.

Following the same procedure as  for $R(D^{(*)})$ ratios, it is straight
forward to find the pQCD predictions for the new $R_{\rm D}^{\rm l,\tau}$ numerically:
$R_{\rm D}^{\rm l} = 0.450^{+0.064}_{-0.051}$ and $R_{\rm D}^{\rm \tau} = 0.642^{+0.081}_{-0.070}$,
here the dominant errors come from the uncertainty of
$\omega_{\rm B}=0.40\pm 0.04$ GeV and $m_{\rm c}=1.35\pm 0.03$ GeV.
The error of the pQCD predictions for ratio $R_{\rm D}^{\rm l}$ and $R_{\rm D}^{\rm \tau}$ is about $15\%$.

\section*{5. Summary} \label{sec:5}

In summary, we studied the semileptonic decays
$B \to D^{(*)} l^- \bar{\nu}_{\rm l}$ in the framework of the SM
by employing the pQCD factorization approach.
From the numerical calculations and phenomenological analysis we found that
\begin{itemize}
\item[1)]
The pQCD predictions for the branching ratios
${\cal B}(B \to D^{(*)} l^-\bar{\nu}_{\rm l})$ agree well with  other SM predictions
and the measured values within one standard deviation.

\item[2)]
For the isospin-constrained ratios $R(D)$ and $R(D^*)$, the pQCD predictions are
\beq
R(D)=0.430^{+0.021}_{-0.026}, \quad R(D^*)=0.301 \pm 0.013. \label{eq:rdrds}
\eeq
We therefore provide a SM interpretation for the BaBar's $R(D^{(*)})$ anomaly.

\item[3)]
For the newly defined ratios $R_{\rm D}^{\rm l,\tau}$, the pQCD predictions are
\beq
R_{\rm D}^{\rm l} = 0.450^{+0.064}_{-0.051}, \quad
R_{\rm D}^{\rm \tau} =0.642^{+0.081}_{-0.070}, \label{eq:rdlrdtau}
\eeq
These new ratios may be more sensitive to the QCD dynamics of the considered
decays than the ratios $R(D^{(*)})$, we therefore
suggest the experimental measurements for them in the forthcoming experiments.

\end{itemize}

\begin{acknowledgments}
We wish to thank Hsiang-nan Li, M.F. Sevilla, Wei Wang and Xin Liu
for valuable discussions.
This work was supported by the National Natural Science Foundation of
China under Grant No.10975074 and 11235005.

\end{acknowledgments}



\end{document}